\documentstyle[pre,aps]{revtex}
\tightenlines
\begin{document}
\title{Time Reversal and Exceptional Points } 
\author{H.L.~Harney$^1$ and W.D.~Heiss$^{1,2}$}
\address{$^1$Max-Planck-Institut f\"ur Kernphysik, 69029 Heidelberg,
             Germany\\
         $^2$Department of Physics, University of Stellenbosch, 
             7602 Matieland, South Africa}  
\date{\today}
\maketitle
\begin{abstract}  
Eigenvectors of decaying quantum systems are studied at exceptional
points of the Hamiltonian. Special attention is paid to the
properties of the system under time reversal symmetry breaking.
At the exceptional point the chiral character of the system ---
found for time reversal symmetry --- generically persists. It is,
however, no longer circular but rather elliptic. 
\end{abstract}
\medskip

In a system described by a non-hermitian Hamiltonian $H$,
a surprising phenomenon can occur: the coalescence of two
eigenmodes. This means that two eigenvalues merge such that
there is only one eigenvector. As a consequence, $H$ cannot be
diagonalized by a similarity transformation.
Considering $H$ to depend on
some parameter $\lambda\, ,$ the value $\lambda_c\, ,$
where this happens, is called  an exceptional point 
(EP)\cite{Kato}.  It is well known that hermitian operators cannot
have any EPs: at a degeneracy of two of their eigenvalues, the space
of eigenvectors is two-dimensional. 

In the vicinity of an EP, 
the eigenvalues \emph{and} eigenvectors show branch point
singularities \cite{Hesa,hei,BerryEP,Mondragon} as functions of
$\lambda\, .$ This contrasts with a two-fold degeneracy in a hermitian
matrix, where no singularity but rather a diabolic point 
\cite{BerryDP} occurs. EPs have been observed in laser induced
ionization of atoms \cite{Latinne}, in acoustical systems
\cite{Shuva}, microwave cavities \cite{BrentanoEP,DemboEP}, in
optical properties of certain absorptive media \cite{Pancha,BerryPancha,bede},
and in ``crystals of light'' \cite{Oberthaler}. Models for Stark 
resonances in atomic physics have been analysed in terms of EPs and 
their connection to diabolic points discussed \cite{kors}.
The broad variety of
physical systems showing EPs indicates that their occurrence is
generic. 

The observation of EPs is possible in decaying quantum systems.
So far only complex symmetric Hamiltonians have been considered. 
Such ``effective'' Hamiltonians are used to model
decaying or resonant systems when invariance under time reversal
prevails. They are obtained by eliminating open decay channels 
from explicit consideration. Therefore, the possibility for the 
system to decay is not at variance with time reversal symmetry.

For complex symmetric $H\, $, a recent theoretical paper 
\cite{heha} has found the eigenfunction
$|\psi_{\rm EP}\rangle$ at the EP to be of the form
\begin{equation}
|\psi_{\rm EP}\rangle =\pmatrix{ \pm i \cr 1} .
                                                  \label{wep}
\end{equation}
The phase difference
$\pm i$ between the components of the state vector is independent of an 
arbitrary two-dimensional orthogonal transformation \cite{heha}; in fact, the 
$|\psi_{\rm EP}\rangle$ are eigenstates of an orthogonal transformation
and are therefore independent of a particular choice of the basis.
The phase $i$ has been confirmed experimentally \cite{dembph}.

In the present note, we again address the eigenfunction at the EP
--- for a situation, where time reversal symmetry is broken. This
does not necessarily mean that it is broken on a fundamental level
as it is in the system of the neutral $K$-mesons. An external magnetic 
field applied to a moving charge provides time reversal symmetry breaking.
Similarly, one can introduce magnetic elements into microwave cavities
that allow only one direction for a traveling wave \cite{ach}.

In the vicinity of an EP, where two (and only two) eigenvalues merge,
an $n$-dimensional system can locally be represented by a two-state
system \cite{heha}. Therefore we confine ourselves to two-dimensional 
$H$ in the sequel. 

Let $H$ be the sum
\begin{equation}
H=H_0+\lambda H_1
              \label{hamt}
\end{equation}
of two hermitian operators $H_0,H_1$ with $H_1$ multiplied by a
complex strength parameter $\lambda\, .$  In order that there be
exceptional points $\lambda_c$, the operators $H_0$ and $H_1$ must
not commute. If they do, $H$ can be diagonalized for every complex 
$\lambda$. We write
\begin{equation}
H_0=U_0\epsilon U_0^{\dagger}
\quad\quad {\rm and}\quad\quad
H_1=U_1\omega U_1^{\dagger}\, ,
                                    \label{3}
\end{equation}
where
\begin{equation}
\epsilon =\pmatrix{\epsilon_1 & 0 \cr 0 & \epsilon_2}
\quad\quad {\rm  and}\quad\quad
\omega   =\pmatrix{\omega_1 & 0 \cr 0 & \omega_2}\, ,
                                        \label{4}
\end{equation}
and $U_0, U_1$ are unitary matrices. Throughout the present paper
we assume that $\epsilon$ and $\omega$ are different from a 
multiple of the unit matrix. Then $H_0, H_1$ do not  commute if 
$U_0, U_1$ are different from each other. 

The $U_k$ can be parameterized by two angles $\phi$ and $\tau$ 
such that
\begin{equation}
U_k=U(\phi_k,\tau_k)
\quad\quad {\rm where}\quad\quad
  k=0,1
                     \label{5}
\end{equation}
and
\begin{equation}
U(\phi ,\tau )
   =\pmatrix{\cos\phi             &-\sin\phi\exp (i\tau)\cr
             \sin\phi\exp (-i\tau)&\cos\phi}\, .
                                     \label{unit}
\end{equation}
A general unitary transformation $U_g$ in two dimensions actually
has four parameters. It can be represented as
\begin{equation}
U_g=U\pmatrix{\exp (i\gamma_1)&0               \cr
              0               &\exp (i\gamma_2)
             }\, .
                                  \label{7}
\end{equation}
Since $H_k$ is independent of the phases $\gamma_1,\gamma_2\, ,$
we may set them equal to zero. Thus, (\ref{unit}) is the
most general unitary transformation in the context of
Eq. (\ref{3}). 

The complex strength parameter $\lambda$ allows for the system 
to decay. Since we assume that the time reversal operator 
$\cal T$ equals the complex conjugation $K$, the Hamiltonian 
$H$ is time reversal symmetry breaking if $\tau_0\neq 0$ 
or $\tau_1\neq 0$.

Let us discuss the consequences of this model in two steps: First, 
the simplification $\tau_1=\tau_2$ is introduced, and 
second, the general case is discussed.

{\it A special case of time reversal symmetry breaking.}
We assume that $\tau_0$ and $\tau_1$ equal each other,
i.e. 
\begin{eqnarray}
\tau &=&\tau_0\nonumber\\
     &=&\tau_1\, .
                 \label{tau}
\end{eqnarray}

At the EPs, the eigenvectors are
\begin{equation}
|\psi^{\pm }_{{\rm EP}}\rangle
   \propto\pmatrix{\pm ie^{i\tau}\cr 1}\, ,
                                  \label{wfph}
\end{equation}
respectively. One recognizes this by using the time reversal
symmetric result (\ref{wep}) together with the observation 
that the present $H$ can be written in the form
\begin{equation}
H=z(\tau)\left(U(\phi_0,0)\epsilon U^{\dagger}(\phi_0,0)
               +\lambda U(\phi_1,0)\omega U^{\dagger}(\phi_1,0)
         \right)z^{\dagger}(\tau )\, ,
                                    \label{Htau}
\end{equation}
where $z$ is the matrix
\begin{equation}
z(\tau )=\pmatrix{\exp ({\rm i}\tau /2) & 0 \cr
                  0                     & \exp (-{\rm i}\tau /2)
                 }\, .
                          \label{z}
\end{equation}
The result (\ref{wfph}) differs from Eq. (10) of \cite{heha} by the
phase $\tau$. For $\tau =0$, one retrieves the time
reversal invariant situation described there. 

The left hand eigenvectors $\langle\tilde{\psi}^{\pm}_{\rm EP}|$ at
the EPs are
\begin{equation}
\langle\tilde{\psi}^{\pm}_{\rm EP}|=(\pm ie^{-i\tau},\, 1)\, .
                                \label{wfph2}
\end{equation}
This differs from Eq. (5) of \cite{heha} in that
$|\tilde{\psi}^{\pm}_{\rm EP}\rangle$ is not the complex conjugate
of the right hand eigenvector $|\psi^{\pm}_{\rm EP}\rangle\, .$
Note, however, that relation (9) of \cite{heha} persists: 
the inner product of the left and right hand eigenvectors vanishes, {\it viz.}
\begin{equation}
\langle\tilde{\psi}^{\pm}_{\rm EP}|\psi^{\pm}_{\rm EP}\rangle =0\, .
                                 \label{scal}
\end{equation}
Therefore the biorthogonal normalization is impossible at the EP.

The present case covers the  even more special situation where
$\phi_0=0$, whence
\begin{equation}
H_0=\epsilon\, .
      \label{H0diag}
\end{equation}
The evaluation of this case provides the basis for the later
evaluation of the general case.
If (\ref{H0diag}) holds, the eigenvalues are
\begin{equation}
E_{1,2}={1\over 2}
        \bigg(\epsilon_1+\epsilon_2+\lambda (\omega_1+\omega_2)
        \bigg)\pm R\, ,
                                \label{eig}
\end{equation}
where
\begin{equation}
R={1\over 2}\sqrt{
  (\epsilon_1-\epsilon_2)^2
        +\lambda^2(\omega_1-\omega_2)^2
        +2\lambda (\epsilon_1-\epsilon_2)(\omega_1-\omega_2)\cos 2\phi_1}
                              \, .
                                \label{dis}
\end{equation}
The two levels coalesce at $R=0\, .$ This yields two EPs at
\begin{equation}
\lambda ^{\pm }_c=-{\epsilon_1-\epsilon_2\over 
                             \omega_1-\omega_2
                            }e^{\pm 2i\phi_1}.
                                  \label{ep}
\end{equation}
Note that by our assumptions $\epsilon_1\neq\epsilon_2,\; \omega_1\neq\omega_2$ 
and $\phi_1\neq 0$. 

For a given $\tau$, the transformations
$U(\phi,\tau )$ with $-\pi\leq\phi <\pi$ form a subgroup of the 
unitary matrices. The eigenvectors (\ref{wfph}) are invariant 
under the transformations of
this group because $|\psi^{\pm}_{\rm EP}\rangle$ is eigenvector of 
every element of the group, i.e.
\begin{equation}
U(\phi,\tau )|\psi^{\pm}_{\rm EP}\rangle
   =\exp (\pm i\phi)|\psi^{\pm}_{\rm EP}\rangle\, .
                               \label{eivec}
\end{equation}
This parallels the result stated above and in \cite{heha}  in that 
the vector given in Eq. (\ref{wep}) is an eigenvector of all 
orthogonal transformations. In the more general
cases of time reversal symmetry breaking there is no longer such 
symmetry.

{\it The general case of time reversal symmetry breaking.}
We assume that both, $\phi_0$ and $\phi_1$, are different from zero
and $\tau_0\neq\tau_1$.

One can, of course, transform to the eigenbasis of $H_0\, .$ 
Then $H$ takes the diagonal form used above. 
However, $\cal T$ is no longer equal to $K$, it rather is 
${\cal T} = U^{\dagger }(\phi_0,\tau_0)KU(\phi_0,\tau_0)$.
By this change of the basis, we obtain the general form of the eigenfunction 
at the EP.

The diagonalisation of $H_0$ brings $H$ into the form
\begin{eqnarray} 
\tilde H &=& \epsilon
              +\lambda U^{\dagger }(\phi_0,\tau_0)H_1U(\phi_0,\tau_0)
                                                          \nonumber\\
         &=& \epsilon
              +\lambda U^{\dagger }(\phi_0,\tau_0)U(\phi_1,\tau_1)
                       \omega
                       U^{\dagger }(\phi_1,\tau_1)U(\phi_0,\tau_0)\, .
                                                         \label{Htil}
\end{eqnarray}
According to the above discussion, the eigenfunction at the EP is given by
\begin{equation}
|\psi^{\pm }_{{\rm EP}}\rangle _{\tilde H}
   \propto\pmatrix{\pm ie^{i\xi}\cr 1}\, ,
                                  \label{wfphtil}
\end{equation}
where the phase $\xi $ and the position of the EP are functions of
$\phi_0,\tau_0,\phi_1,\tau_1$. The explicit dependence is deferred to the 
appendix. As a result, the state vector of the Hamiltonian $H$ at the
EP is given by
\begin{equation}
|\psi^{\pm }_{{\rm EP}}\rangle _H
   \propto U(\phi_0,\tau_0)\pmatrix{\pm ie^{i\xi}\cr 1}
\end{equation}
which may be rewritten as
\begin{equation}
|\psi^{\pm }_{{\rm EP}}\rangle _H
   \propto\pmatrix{\pm i e^{-i\xi }
                   \left(e^{2i\xi}\cos^2\phi_0+e^{2i\tau_0}\sin^2\phi_0
                   \right) \cr 
                   1\pm \sin(2\phi _0)\sin(\tau_0-\xi)
                  }\, . 
                                                  \label{wffin}
\end{equation}
Here, the lower component has been chosen real. Note that the scalar product
of the left hand and right hand eigenvector again vanishes as in (\ref{scal}).

We discuss this result. (i) For $\phi_0=0$, corresponding to 
diagonal $H_0$, the form of Eq. (\ref{wfph}) 
is retrieved from (\ref{wffin}).
(ii) For $\tau _0=0$, corresponding to time reversal invariant $H_0$,
the upper component of (\ref{wffin}) simplifies to 
$\pm i\cos(\xi )\mp \cos(2\phi _0)\sin\xi$. Even
in this case, the ratio between the upper and lower component is not 
a phase factor but can assume any value in the complex plane. 
Of course, large values of the ratio correspond to small values of
the lower component which can in fact vanish. 
(iii) If $\tau_0=\tau_1$ while maintaining $\phi _0\ne \phi _1$, 
Eq. (\ref{wffin}) leads back to Eq. (\ref{wfph}).

{\it In conclusion}: the ``universal'' phase of $\pi /2$ in the 
eigenvector at an EP is no longer universal if time reversal symmetry
is broken. The phase as well as the relative amplitude in the
eigenvector at the EP can be manipulated in an experiment that allows
to control the violation of this symmetry. 

If --- in analogy with wave optics \cite{bede} and as was done in 
\cite{dembph} --- one associates a circularly
polarized wave with the eigenvector (\ref{wep}), time reversal
symmetry breaking leads to the elliptically polarized wave
(\ref{wffin}). An elliptical wave can be generated in two ways: 
the phase of the upper component in (\ref{wffin}) is different from
$\pm\pi /2$ or the amplitude ratio is not unity. The limit of a
linearly polarized wave is obtained when the upper component is real. 
If this happens, the chiral character of the wave function at an EP 
is lost. We emphasize, 
however, that even this dramatic effect of time reversal symmetry 
breaking upon the wave function does not alter the fact that 
only one mode can occur at the EP. Time reversal symmetry breaking 
simply changes the amplitude ratio such that it may assume 
any complex value. This includes in particular the special cases
where either the upper or the lower component in (\ref{wffin}) vanishes.

We believe that these results can be verified in 
future experiments with quantum dots in a magnetic
field and with microwave cavities containing suitable magnetic
elements \cite{ach}.

{\bf Acknowledgement} WDH acknowledges the warm hospitality of the
theory group at the Max-Planck-Institute for Nuclear Physics at
Heidelberg. Both authors acknowledge stimulating discussions with 
their experimental colleagues A.~Richter, H.-D. Gr\"af, and 
C.~Dembowski at the Institute for Nuclear Physics of the TU Darmstadt.

\appendix
\section{The phase $\xi$}
\label{app}
The product 
\begin{equation}
U_0^{\dagger}U_1=U^{\dagger}(\phi_0,\tau_0)U(\phi_1,\tau_1)
                                             \label{app1}
\end{equation}
is equal to
\begin{equation}
U_0^{\dagger}U_1
  = \pmatrix{\cos\phi_0\cos\phi_1+\sin\phi_0\sin\phi_1
                                    e^{i(\tau_0-\tau_1)} &
              \cos\phi_1\sin\phi_0e^{i\tau_0}
              -\cos\phi_0\sin\phi_1e^{i\tau_1} \cr
              -\cos\phi_1\sin\phi_0e^{-i\tau_0}
              +\cos\phi_0\sin\phi_1e^{-i\tau_1} &
              \cos\phi_0\cos\phi_1
              +\sin\phi_0\sin\phi_1e^{-i(\tau_0-\tau_1)}  
             }\, .
                                            \label{app2}
\end{equation}
We introduce the phase of the (1,1)-element, viz.
\begin{equation}
\gamma=\arg\left(\cos\phi_0\cos\phi_1
                 +\sin\phi_0\sin\phi_1e^{i(\tau_0-\tau_1)}
           \right)\, ,
                                      \label{app3}
\end{equation}
and write (\ref{app1}) in the form
\begin{equation}
U_0^{\dagger}U_1
   =\pmatrix{\cos\beta  & -\sin\beta e^{i\xi} \cr
                       \sin\beta e^{-i\xi }  & \cos\beta 
                      }z(2\gamma)\, .
                                    \label{app4}
\end{equation}
Here, $z$ is the matrix defined in (\ref{z}), 
and $\beta$ is given by
\begin{equation}
\cos\beta 
   =\left(\cos^2\phi_0\cos^2\phi_1
           +\sin^2\phi_0\sin^2\phi_1
           +2\cos\phi_0\cos\phi_1\sin\phi_0\sin\phi_1\cos(\tau_0-\tau_1)
    \right)^{1/2}
                                                 \label{app5}
\end{equation}
and $\xi$ by
\begin{equation}
\xi=\arg\left(\cos\phi_1\sin\phi_0e^{i\tau_0}
              -\cos\phi_0\sin\phi_1e^{i\tau_1}
        \right)+\gamma\, .
                             \label{app6}
\end{equation}
The exceptional point occurs at
\begin{equation}
\lambda_{\rm EP}=-{\epsilon_1-\epsilon_2\over \omega_1-\omega_2}
                  e^{\pm 2i\beta}\, .
                               \label{app7}
\end{equation}

\end{document}